\documentclass[12pt]{article}
\pdfoutput=1
\usepackage{hyperref}
\usepackage{graphicx}
\usepackage{graphics}
\usepackage{dsfont}
\usepackage{epsfig}
\usepackage{amsmath,amssymb,amsthm,amscd}

\newcommand{\Res}{\textrm{Res}}

\numberwithin{equation}{section}

\newcommand{\be}{\begin{equation}}
\newcommand{\ee}{\end{equation}}
\newcommand{\ba}{\begin{aligned}}
\newcommand{\ea}{\end{aligned}}

\begin{document}
\begin{titlepage}
{}~ \hfill\vbox{ \hbox{} }\break

\rightline{USTC-ICTS/PCFT-20-16}

\vskip 3 cm

\centerline{\Large 
\bf  
Note on Quantum Periods } 
\vskip 0.2 cm
\centerline{\Large 
\bf 
 and a TBA-like System}

\vskip 0.5 cm

\renewcommand{\thefootnote}{\fnsymbol{footnote}}
\vskip 30pt \centerline{ {\large \rm Min-xin Huang
\footnote{minxin@ustc.edu.cn}  } } \vskip .5cm  \vskip 20pt 

\begin{center}
{Interdisciplinary Center for Theoretical Study,  \\ \vskip 0.1cm  University of Science and Technology of China,  Hefei, Anhui 230026, China} 
 \\ \vskip 0.3 cm
{Peng Huanwu Center for Fundamental Theory,  \\ \vskip 0.1cm  Hefei, Anhui 230026, China} 
\end{center}

\setcounter{footnote}{0}
\renewcommand{\thefootnote}{\arabic{footnote}}
\vskip 60pt
\begin{abstract}
There is an interesting relation between the quantum periods on a certain limit of local $\mathbb{P}^1\times \mathbb{P}^1$ Calabi-Yau space and a TBA (Thermodynamic Bethe Ansatz) system appeared in the studies of  ABJM (Aharony-Bergman-Jafferis-Maldacena) theory. We propose a one-parameter generalization of the relation. Furthermore, we derive the differential operators for quantum periods and the TBA-like equation in various limits of the generalized relation.

\end{abstract}

\end{titlepage}
\vfill \eject


\newpage

\baselineskip=16pt

\tableofcontents

\section{Introduction and Summary}

In geometry, we often compute period integrals, which are the integrals of differential forms in certain cohomological classes over cycles of the geometry. For Calabi-Yau three-folds, the classical periods of the holomorphic 3-form over 3-cycles play important roles in mirror symmetry \cite{Candelas:1990}. The classical periods in mirror symmetry satisfy a set of differential equations known as the Picard-Fuchs equations. Basically, because the cohomology space is finite dimensional, we can construct Picard-Fuchs operators by linear combinations of derivatives of complex structure moduli, whose actions on the differential form are exact forms, so that the integrals vanish over cycle. 

There have been many works to generalize the notion to quantum geometry and quantum periods, see e.g. \cite{Aganagic}. In the context of mirror symmetry, we consider local Calabi-Yau geometries which can be described by complex one-dimensional curves. Quantization of the geometry amounts to promoting the complex coordinates of the curves to canonical conjugate operators $[\hat{x}, \hat{p}] =i\hbar$. For notational convenience with the $i$ factor we also denote $\epsilon \equiv i\hbar$. The wave function of the quantum system has a standard WKB expansion $\psi(x) = \exp [ \frac{1}{\epsilon} \int^x w(x^\prime) d x^\prime]$, where the integrand in the exponent has a power series expansion in $\epsilon$ parameter. The quantum periods can be computed by integrals of  $w(x)dx$ over cycles, which in this case are contour over the complex $x$ plane. In the classical limit $\epsilon\rightarrow 0$, this reduces the classical periods, which are integrals of the canonical differential one-form $pdx$ over cycles. An important property of the quantum periods is that the higher order contributions can be computed by certain differential operators acting on the classical periods \cite{Mironov:2009, Huang:2012, HKRS}.

The classical periods provide a solution for the prepotential of the Calabi-Yau geometry. The free energy of topological string theory also includes higher genus contributions. More generally, motivated by Nekrasov's calculations \cite{Nekrasov:2002} of instanton partition function of Seiberg-Witten theory, one can define refined topological string theory which has expansion over two small parameters $\epsilon_{1,2}$ \cite{Iqbal:2007}. The conventional genus expansion corresponds to $\epsilon_1+\epsilon_2=0$, while another special limit of setting one parameter to zero e.g. $\epsilon_2=0$ is known as the Nekrasov-Shatashvili (NS) limit \cite{Nekrasov:2009}. One application of the quantum periods is that they can compute the topological free energy in NS limit, in the same way as the classical periods compute the prepotential. This has been studied in Seiberg-Witten theories as well as topological string theory \cite{Mironov:2009, Huang:2012}. Furthermore, exact quantization conditions for quantum systems of mirror curves including novel non-perturbative contributions are conjectured in \cite{Huang:2014, Grassi, Wang:2015}. 

In this note, we consider the case of a well studied local $\mathbb{P}^1 \times \mathbb{P}^1$ Calabi-Yau space, which in a special limit is related to the computations of partition functions of ABJM theory on 3-sphere \cite{ABJM, Kapustin:2009}. On the other hand, the partition function of ABJM theory can be also formulated in terms of fermion gas, and is related to a TBA system, which are studied in many papers \cite{Marino:2011eh, Calvo:2012du, Hatsuda:2012hm, Putrov:2012zi, Hatsuda:2012dt, Okuyama:2015pzt, Furukawa:2019piy}, based on the early seminal papers \cite{Zamolodchikov:1994uw, Tracy:1995ax}. 

In particular, a novel relation between the quantum periods and the TBA system was conjectured and later derived in \cite{Hatsuda:2013oxa, Kallen:2013qla}. The derivation uses ABJM theory as well as its many sophisticated technical ingredients. However, in order to have a deeper understanding as well as exploring possible generalizations to more Calabi-Yau spaces, it is worthwhile to directly study the relation which by itself can be formulated independently without ABJM theory.  

We should note that in our study we will only use a TBA-like difference equation and will not need integral equations which may be more familiar for obtaining the analytic properties of the underlying quantum spectral problem.  We will not study the spectral theory but only solve the difference equations perturbatively around a small complex structure deformation parameter for the purpose of comparing with quantum A-periods. As in the examples in a subsequent paper \cite{Du:2020nwl}, our difference equations are similar to, but are not necessarily equivalent to those of an actual TBA system with good analytic properties. In this sense, this type of difference equations is usually called TBA-like.   

The paper is organized as the followings. In section \ref{secgeneral}, we introduce the notations and propose a one-parameter generalization of the relation between quantum periods and TBA system in \cite{Hatsuda:2013oxa}. In sections \ref{secTBA} and \ref{secPeriods}, we then take another perspective of the relation by expanding perturbatively for small $\epsilon$ parameter, but keep the coefficients exactly as a function of $z$ in terms of differential operators.  We will compute the differential operators to the first few orders for various limits in the generalized setting, and verify that they are the same for the TBA system and the quantum periods.

Subsequently after this paper appeared, we further generalize the relation for a large class of Calabi-Yau geometries \cite{Du:2020nwl}. This later work indicates that the relation studied here is not an isolated phenomenon, but rather universal.

\section{A one-parameter generalization}  \label{secgeneral}  

The mirror curve of the local $\mathbb{P}^1 \times \mathbb{P}^1$ Calabi-Yau model is 
\begin{eqnarray}
e^x + e^p +z_1 e^{-x} + z_2 e^{-p} =1,
\end{eqnarray} 
where $x,p$ are the complex coordinates and $z_1, z_2$ are the complex structure moduli parameters. We promote the complex coordinates to canonical operators $\hat{x} = x, \hat{p} = \epsilon \partial_x$.  The mirror curve acts on a wave function $\psi(x)$ so that 
\begin{eqnarray}
(-1 + e^x + z_1 e^{-x} ) \psi(x) + \psi(x+\epsilon) +z_2 \psi(x-\epsilon) =0 . 
\end{eqnarray} 
It is convenient to use exponential variables $q=e^{\epsilon}, X=e^x$.  A particular choice of $z_1\rightarrow q^{\frac{1}{2}}z, z_2\rightarrow q^{-\frac{1}{2}} z$ corresponds to the calculations of ABJM partition function. Here we will instead consider a one-parameter generalization of the ABJM setting, by keeping the general $\mathbb{P}^1 \times \mathbb{P}^1$ geometry. We introduce the following parametrization with an additional $m$ parameter 
\begin{eqnarray}
z_1 = e^m z, ~~~~ z_2 =e^{-m} z. 
\end{eqnarray} 
The original conjecture in \cite{Hatsuda:2013oxa} corresponds to a special choice $m\rightarrow \frac{\epsilon}{2}$ in our set up. 

One also introduces a function $V(X =e^x) = \frac{\psi(x+\epsilon)}{\psi(x)}$. Instead of the standard WKB expansion of the wave function $\psi(x) = \exp [ \frac{1}{\epsilon} \int^x w(x^\prime) d x^\prime]$, we consider the logarithmic  function $\log V(X) = \frac{1}{\epsilon} \int_x^{x+\epsilon}   w(x^\prime) d x^\prime $, which may differ with $ w(x)$ only by some total derivatives. So their residue is actually the same.  We have a function of four parameters $V(X, \epsilon, z, m)$ which satisfies the equation
\begin{eqnarray} \label{Vequation}
-1+X+\frac{e^m z}{X} + V(X, \epsilon, z, m) +\frac{e^{-m} z}{V(q^{-1} X, \epsilon, z, m) } =0 
\end{eqnarray} 
We can compute the function $V(X, \epsilon, z, m)$ recursively as a perturbative expansion of small $z$. The equation (\ref{Vequation}) is quadratic in the classical limit $\epsilon\rightarrow 0$. Without loss of generality, we choose one of the two solutions which is a nonzero constant in the $z\rightarrow 0$ limit. In this case, the equation becomes linear at $z=0$ with a unique solution $V(X)=1-X$. We take this as the initial function for the small $z$ perturbation, then the first few terms are  
\begin{eqnarray} \label{Vexpansion1.4}
V(X, \epsilon, z, m) = 1 - X - (\frac{e^{-m} q}{q - X} + \frac{e^m}{X} ) z   - (\frac{e^{-2 m} q}
{q^2 - X} + \frac{1}{X}) \frac{q^3  z^2 }{(q - X)^2}  + \mathcal{O}(z^3) .  
\end{eqnarray}

The  function $\log(V)$ has only a finite number of poles in the complex $x$ plane in the small $z$ perturbation. The quantum A-period can be computed by a contour integral of $\log(V)dx$ around a large circle  enclosing all the poles, and by contour deformation the calculation becomes a residue around $x\sim -\infty$. Instead of computing residue of $x$, it is more convenient here to compute equivalently the residue of $X=e^x$ around $X\sim 0$ with an extra factor of $\frac{1}{X}$ since $dx=\frac{dX}{X}$, which would be just the constant term in the Laurent expansion around $X\sim 0$. It also turns out that the residue of $X$ missed a  $\log(z)$ term due to the change of integration variable at infinity. Taking this into account and use the proper combination, we have the quantum A-period 
\begin{eqnarray} \label{quantumAP}
\Pi_A &=&  \log(z) - \Res_{X=0}\frac{2}{X} \log(V(X, \epsilon, z, m)  \\ \nonumber 
&=& \log(z) + 2( e^m + e^{-m}) z + [3 (e^{2 m} +  e^{-2 m}) + 2 (4 + \frac{1}{q} + q)] z^2   
+\mathcal{O}(z^3) ,
\end{eqnarray} 
where  the $\log(z)$ term is well known to be present in the classical periods and responsible for their monodromy. The proper combination of $\log(z)$  with the residue can be fixed by comparing with classical period at $q=1$ which is determined by a Picard-Fuchs differential equation.

On the other hand, the TBA-like system corresponding to the ABJM theory of the choice $z_1\rightarrow q^{\frac{1}{2}}z, z_2\rightarrow q^{-\frac{1}{2}} z$ is described by a function $\eta(X, \epsilon, z)$ of the three variables, which also satisfies a difference equation 
\begin{eqnarray} \label{ABJM2.7}
1 +z [ \eta(qX)+ \eta(X)][\eta(q^{-1}X)+\eta(X) ](X+X^{-1} + q^{\frac{1}{2}} + q^{-\frac{1}{2}} ) =  \eta(X)^2. 
\end{eqnarray} 
It is reasonable to expect that there should be a one-parameter deformation which corresponds to the general local $\mathbb{P}^1 \times \mathbb{P}^1$ Calabi-Yau model described above. After some guessworks, we find such a deformation, which is to simply replace the $q^{\frac{1}{2}} + q^{-\frac{1}{2}}$ term with $e^m +e^{-m}$, where $m$ is  the extra deformation mass parameter. In the generalized model, we now have a four-parameter function $\eta(X, \epsilon, z, m)$, defined by a deformed equation 
\begin{eqnarray}    \label{etaequation}
1 +z [ \eta(qX)+ \eta(X)][\eta(q^{-1}X)+\eta(X) ](X+X^{-1} + e^m +e^{-m}  ) =  \eta(X)^2. 
\end{eqnarray} 

Subsequently after this paper appeared, the above equation (\ref{etaequation}) is derived using the spectral theory in \cite{Du:2020nwl}. We provide some details of the general procedure in Appendix \ref{appendixA}. In our case, the function $u(x)$ in (\ref{TBAgeneral}) can be expressed in terms of Faddeev's quantum dilogarithm function. Using some functional relations one can show that in our case the general equation (\ref{TBAgeneral}) becomes (\ref{etaequation}) with only elementary functions.

One can also solve the function $\eta(X, \epsilon, z, m)$ recursively as a perturbative series of $z$. With the choice of plus sign for the leading term, the first few terms are 
\begin{eqnarray} 
 \eta(X, \epsilon, z, m) &=& 1 + 2 (e^m + e^{-m} + X + X^{-1} ) z + 
2 ( e^m+ e^{- m} +  X + X^{-1}  ) \nonumber \\
&&\cdot [3 (e^m + e^{-m}) +   ( q + 1+q^{-1}) (X + X^{-1})]   z^2 + \mathcal{O}(z^3) .  
\end{eqnarray} 
The relation between quantum A-period and the TBA-like system is then
\begin{eqnarray} \label{relation1.8}
 \Res_{X=0}\frac{1}{X}  \eta(X, \epsilon, z, m) = \theta_z  \Pi_A  (\epsilon, z, m) , 
\end{eqnarray} 
where the differential operator is defined $\theta_z \equiv z\partial_z$.  Although the equations for quantum periods (\ref{Vequation}) and for the TBA-like system ({\ref{etaequation}) look quite different and the expansions are also different, after taking residue, one can check the relation (\ref{relation1.8}) perturbatively for small $z$ where the coefficients are rational functions of $q$ and $e^m$.  Thus we have provided a generalization of the relation in \cite{Hatsuda:2013oxa}. 

We do not provide a rigorous proof of the relation (\ref{relation1.8}). In the ABJM case of TBA-like equation (\ref{ABJM2.7}), a somewhat rigorous derivation is given \cite{Kallen:2013qla}. However, the approach seems very specific to ABJM theory and it is not clear whether the derivation can be conveniently applied to our case of general mass parameter. In lights of the much more general results in our subsequent work \cite{Du:2020nwl}, it is probably preferable to search for a more universal approach for rigorous proof. We may return to this question in future works.

\section{Differential operators for a TBA-like system} \label{secTBA}

We now consider a different expansion, by solving the equations as a perturbative series of $\epsilon$. We can treat the extra parameter $m$ in two ways, either as an independent finite parameter, or it can depend also on $\epsilon$, e.g. scaling like $m = \tilde{m}\epsilon$ with $\tilde{m}$ finite. We will compute in both cases. 

In this section, we study the TBA-like equation ({\ref{etaequation}). First we consider the case of $m$ as an independent finite parameter. We denote the perturbative series and the residue as 
\begin{eqnarray} 
 \eta(X, \epsilon, z, m) =\sum_{n=0}^\infty  \eta_n(X, z, m) \epsilon^n , ~~~~ p_n(z, m) \equiv    \Res_{X=0}\frac{1}{X}  \eta_n(X, z, m) . 
\end{eqnarray}  
Since the equation is invariant under the sign switch $\epsilon\rightarrow -\epsilon$, the coefficients vanish $\eta_n(X, z)=0$ for odd integers $n$. So we only need to consider even terms.

The leading term can be solved by a simple quadratic equation. Our convention uses the solution with plus sign 
\begin{eqnarray}  \label{eta0}
 \eta_0(X, z, m) = \frac{1}{\sqrt{1- 4 (e^m+e^{-m}+X+X^{-1} ) z } }. 
\end{eqnarray} 
We can compute the leading period perturbatively for small $z$ expansion
\begin{eqnarray} \label{p0per}
p_0 &=& 1+2( e^m+e^{-m}) z+6 ( e^{2m}+e^{-2m} +4 ) z^2+20  [e^{3m}+e^{-3m} +9( e^m+e^{-m})] z^3 
\nonumber \\ && +\mathcal{O}(z^4) .
\end{eqnarray}
The exact expression is determined by a Picard-Fuchs differential operator $\mathcal{L}$ so that its action on the leading term  $\mathcal{L} \eta_0(X, z, m)  $  is a total derivative of $x$.  Such operator can not be constructed by linear combinations of $1,\theta_z$ with rational $z$ coefficients, so we need at least up to the second derivative $\theta_z^2$. After some computations we determine the operator 
\begin{eqnarray} \label{PFoperator} 
\mathcal{L} &=& [16(e^m - e^{-m})^2 z^2 -8(e^m + e^{-m}) z +1] \theta_z^2  + 8z [4(e^m - e^{-m})^2 z -e^m-e^{-m} ]\theta_z  \nonumber \\ 
&& + 2z [ 6(e^m - e^{-m})^2 z  -e^m-e^{-m} ].
\end{eqnarray} 
The operator will be compared with the Picard-Fuchs operator of the classical A-period of the local $\mathbb{P}^1\times \mathbb{P}^1$ geometry in the next section. So the leading order period satisfies a second order differential equation $\mathcal{L} p_0 =0$. We note that this operator (\ref{PFoperator}) is uniquely determined up to multiplication from the left by any operator or rational functions of $z$. Otherwise we can make a nonzero linear combination which does not contain $\theta_z^2$.  The coefficient of $\theta_z^2$ term is the discriminant of the curve, and is denoted 
\begin{eqnarray} \label{delta}
\Delta = 16(e^m - e^{-m})^2 z^2 -8(e^m + e^{-m}) z +1. 
\end{eqnarray} 

The higher order terms $ \eta_n(X, z, m)$ can then be written as linear combinations of $\eta_0(X, z, m)$ and 
$\theta_z \eta_0(X, z, m)$, plus a total derivative of $x$. So $p_0$ and $\theta_z p_0$ provide a linear basis for the higher period $p_n$. After some computations, we determine the differential operators for next two orders 
\begin{eqnarray} \label{p2operator}
p_2 &=&  [16(e^m+e^{-m})(e^m-e^{-m})^2z^2 - 8 (e^{2m} +e^{-2m} -6) z +e^m+e^{-m} ] \frac{z \theta_z p_0 }{6\Delta}  \nonumber \\ && 
+  [ 4(e^m+e^{-m})(e^m-e^{-m})^2z -e^{2m} -e^{-2m} +10  ]\frac{z^2 p_0}{3\Delta}  ,  \\
p_4 &=&  \{ 2048 (e^m -e^{-m} )^4 [e^{3m}+ e^{-3 m}  - 85 (e^{ m}+ e^{- m})  ] z^6  
-  512 (e^{m} -e^{-m}  )^2 [3(e^{4m} +e^{-4m}  ) \nonumber \\ && + 494 (e^{2m} + e^{-2m}  ) + 3038 ] z^5
     +  2048 [107(e^{3m} + e^{-3m}  ) - 191 (e^{m} +e^{-m}  ) ] z^4 
    \nonumber \\ &&  +  64  [ 5(e^{4m} + e^{-4m}  ) - 658 (e^{2m} + e^{-2m}  ) + 2474 ] z^3  - 
 8  [15(e^{3m} + e^{-3m}  )  \nonumber \\ && + 109 (e^{m} + e^{-m}  ) ]z^2 
 + 2  [9(e^{2m} +e^{-2m}  ) + 292 ] z
- (e^{m} +e^{-m}  ) 
   \} \frac{z \theta_z p_0 }{360 \Delta^3}  \nonumber \\ && 
+  \{    512 (e^{ m}- e^{- m})^4 [(e^{ 3m}+ e^{- 3m}) - 85 (e^{ m}+ e^{- m})]  z^5  - 
 128  (e^{ m} - e^{- m})^2 \nonumber \\ && \cdot  [ 2(e^{ 4m}+ e^{- 4m}) + 519 (e^{ 2m}+ e^{- 2m})+ 2990 ] z^4 - 
 64 [ e^{ 5m}+ e^{- 5m}  - 733(e^{ 3m}+ e^{- 3m}) \nonumber \\ &&  + 1404 (e^{ m}+ e^{- m}) ] z^3 + 
 8 [ 8(e^{ 4m}+ e^{- 4m})  - 907 (e^{ 2m}+ e^{- 2m})  + 3798] z^2  \nonumber \\ &&
 - 2  [ 7(e^{ 3m}+ e^{- 3m}) + 79 (e^{ m}+ e^{- m}) ] z
   + e^{ 2m}+ e^{- 2m}+ 62 
 \} \frac{z^2 p_0}{180\Delta^3} .  \label{p4operator}
\end{eqnarray}  
The calculations here are similar as in the previous papers e.g. in \cite{Huang:2012}. We use a proper ansatz  with some unknown coefficients for the total $x$  derivative which has the same analytic structure as $ \eta_n(X, z, m)$, whose denominator is an odd power of the square root factor in (\ref{eta0}). The square root factor can be cancelled out and then we have a polynomial equation of $X$. Comparing the coefficients of $X$ gives rise to a system of linear equations, with the variables $z, m$ treated as constant parameters and the unknowns also including the coefficients of $p_0$ and $\theta_zp_0$. The solution of the linear equations gives the above expressions (\ref{p2operator}, \ref{p4operator}) which will be compared with those in the calculations of quantum periods in the next section.

Next we study another parametrization $m = \tilde{m}\epsilon$ with $\tilde{m}$ finite, where a choice of  $\tilde{m}=\frac{1}{2}$ corresponds to the ABJM case. In this case we denote the perturbative series and the residue as 
\begin{eqnarray} 
 \eta(X, \epsilon, z,  \tilde{m}\epsilon ) =\sum_{n=0}^\infty  \eta_n(X, z, \tilde{m}) \epsilon^n , ~~~~ p_n(z, \tilde{m}) \equiv    \Res_{X=0}\frac{1}{X}  \eta_n(X, z, \tilde{m} ) . 
\end{eqnarray}

The leading order calculations and the Picard-Fuchs operator are obtained by simply setting $m\rightarrow 0$ in the  above equations (\ref{eta0}, \ref{p0per}, \ref{PFoperator}). The higher order periods are different since the $\epsilon$ dependence in $m$ contributes in the expansion. Again after some computations, we find the differential operators that determine the higher order periods 
\begin{eqnarray}
p_2 &=&  \frac{z}{3(1-16z)} [ 2  (3 \tilde{m}^2 +4 z - 48 \tilde{m}^2 z)p_0 + (1+12 \tilde{m}^2 +16 z-192 \tilde{m}^2 z) \theta_z p_0 ], \nonumber \\ 
p_4 &=&  [15 \tilde{m}^4 - 4 (-8 + 15 \tilde{m}^2 + 45 \tilde{m}^4) z - 
 4 (43 - 1680 \tilde{m}^2 + 2880 \tilde{m}^4) z^2 \nonumber \\ && + 
 320 (25 - 432 \tilde{m}^2 + 816 \tilde{m}^4) z^3 - 6144 (7 - 120 \tilde{m}^2 + 240 \tilde{m}^4) z^4 ]  \frac{z p_0}{90 (1-16z)^3}  \nonumber \\ &&
 +[-1 + 30 \tilde{m}^2 + 60 \tilde{m}^4 + (310 - 1080 \tilde{m}^2) z - 
 32 (31 - 1200 \tilde{m}^2 + 2160 \tilde{m}^4) z^2  \nonumber \\ && + 
 512 (73 - 1260 \tilde{m}^2 + 2400 \tilde{m}^4) z^3 - 
 24576 (7 - 120 \tilde{m}^2 + 240 \tilde{m}^4) z^4] \frac{z \theta_z p_0 }{180(1-16z)^3} . \nonumber \\  \label{operator3.8}
\end{eqnarray}

\section{Differential operators for quantum periods} \label{secPeriods}

In this section, we solve the equation for quantum periods (\ref{Vequation})  as a perturbative series of $\epsilon$ and the results can be compared with those of the previous section \ref{secTBA}.  In practice it turns out the calculations in  this section are much more complicated than those in the previous section using the TBA-like system. In this sense, the relation with the TBA-like system provides a simpler way to compute the quantum periods and their associated differential operators for Calabi-Yau geometries.

Again we first consider the case of $m$ as an independent finite parameter. We denote the perturbative series  as 
\begin{eqnarray} 
\log V(X, \epsilon, z, m) =\sum_{n=0}^\infty  w_n(X, z, m) \epsilon^n . 
\end{eqnarray} 
We expand the quantum A-period (\ref{quantumAP}) as 
\begin{eqnarray}
 \Pi_A  (\epsilon, z, m) = \sum _{n=0}^\infty  \Pi_A^{(n)}  (z, m) \epsilon^n . 
\end{eqnarray} 
Unlike in previous section, the odd $n$ power terms $ w_n(X, z, m)$ do not simply vanish, but are still total derivative of $x$, as familiar in calculations of quantum periods \cite{Huang:2012}. So the terms $ \Pi_A^{(n)}$ with  odd $n$ in the expansion of  quantum A-period vanish. 

In order to later compare with the TBA-like system, we denote the derivative of the quantum period  $ \tilde{p}_n(z, m) = \theta_z  \Pi_A^{(n)}  ( z, m)   $, where we use the tilde symbol to distinguish from the notation of TBA-like system in the previous section, though they are equivalent through our generalized proposal. These coefficients are 
\begin{eqnarray} \label{period2.2} 
\tilde{p}_0(z, m) &=&   \Res_{X=0}\frac{1}{X} [1-2\theta_z w_0(X, z, m) ],  \nonumber \\
  \tilde{p}_n(z, m) & = &  - \Res_{X=0}\frac{2}{X} \theta_z w_n (X, z, m),~~ n\geq 1. 
\end{eqnarray}  

The functions $w_n(X, z, m)$ can be solved recursively. The leading term is solved by a simple quadratic equation. Since we will take residue, we assume $X\sim 0$ and take the branch which have the same leading term as (\ref{Vexpansion1.4}) for $z=0$ 
\begin{eqnarray} 
w_0(X, z, m) = \log[ \frac{1}{2} (1- X-\frac{e^m  z}{X} +\frac{e^{-\frac{m}{2}} }{X} \sqrt{-4 X^2 z+e^m (-X+X^2+e^m z)^2)} ] . \nonumber \\ 
\end{eqnarray} 
The Picard-Fuchs operator is more complicated to derive than the previous section. Since the leading order period $\tilde{p}_0$ should be the same as $p_0$ in the previous section, we have $\mathcal{L} \theta_z  \Pi_A^{(0)}  = \mathcal{L} \tilde{p}_0 = 0$ with the operator $\mathcal{L}$ in (\ref{PFoperator}). So the Picard-Fuchs operator for the classical period is simply $\mathcal{L} \theta_z$, as one can directly verify that $\mathcal{L} \theta_z (\log z - 2w_0)$  is indeed a total derivative of $x=\log(X)$, with no residue around $X\sim 0$ after dividing by X. We note that a constant is also a total derivative but still has non-vanishing residue $\Res_{X=0}\frac{1}{X} =1$, due to the monodromy of $\log(X)$ function. So here the $\log(z)$ term is needed so that the total contribution has no monodromy  in our case. 

We consider the higher order periods. The differential operators for the local $\mathbb{P}^2\times \mathbb{P}^1$ Calabi-Yau model are computed in our previous paper \cite{Huang:2012}. The results are 
\begin{eqnarray}
\Pi_A^{(2)}   &=& -\frac{z_1+z_2}{6} \theta_z \Pi_A^{(0)} + \frac{1-4z_1-4z_2}{12} \theta_z^2 \Pi_A^{(0)}, \nonumber \\
\Pi_A^{(4)} &=& \frac{1}{360\Delta^2} \{ 2[  z_1^2 (1 - 4 z_1)^3 + z_2^2 (1 - 4 z_2)^3+4z_1z_2(8 - 37 z_1  - 37 z_2 - 328 z_1^2  \nonumber \\ && + 1528 z_1 z_2  - 328 z_2^2    + 1392 z_1^3 
-  1376 z_1^2 z_2  - 1376 z_1 z_2^2 + 1392 z_2^3)]\theta_z \Pi_A^{(0)} \nonumber \\ && +[  -z_1 (1 - 4 z_1)^4 - z_2 (1 - 4 z_2)^4  +4z_1z_2(69 - 192 z_1 - 192 z_2 - 1712 z_1^2  \nonumber \\ && + 6880 z_1 z_2 - 1712 z_2^2 + 5568 z_1^3   - 
 5504 z_1^2 z_2  - 5504 z_1 z_2^2 + 5568 z_2^3 )]\theta_z^2 \Pi_A^{(0)} \},  \nonumber 
\end{eqnarray} 
where the discriminant is (\ref{delta}) and the parametrization is $z_1= e^m z, z_2= e^{-m} z$. We act the operator $\theta_z$ on both sides of the equations, and use the Picard-Fuchs operator  $\mathcal{L}$ in (\ref{PFoperator}) to eliminate the second derivatives of $\tilde{p}_0$. In this way, we can derive the differential operators for $\tilde{p}_2, \tilde{p}_4$ as linear combinations of $\tilde{p}_0$ and $\theta_z \tilde{p}_0$. We verify that the results are the same as from the TBA-like system in (\ref{p2operator}, \ref{p4operator}).

Finally we consider the parametrization $m = \tilde{m}\epsilon$ with $\tilde{m}$ finite. However for general  $\tilde{m}$, the computations for higher order periods are quite complicated. Instead of working out the general case, as an illustrative example, we study a special case   $\tilde{m} =\frac{1}{2}$, corresponding to the ABJM theory. We determine the differential operators for $\tilde{p}_2, \tilde{p}_4$ in this special case and verify the results are again the same from the TBA-like system in (\ref{operator3.8}).

Some other well studied local Calabi-Yau models, such as the local $\mathbb{P}^2$ model, also have the same feature that the Picard-Fuchs operator for classical periods can be written as $\mathcal{L} \theta_z$, where $\mathcal{L}$ is a second order differential operator.  The corresponding TBA-like equations are studied in \cite{Okuyama:2015pzt}, and compared with quantum periods in a semiclassical $\hbar$ expansion. It would be interesting study whether this can be tested for finite $\hbar$ as in this paper.  A connection of ABJM theory to  local $\mathbb{P}^2$ model is found by studying the partition functions on ellipsoid \cite{Hatsuda:2016uqa}. It would be interesting to explore whether such a connection can be found for quantum periods as well.

\vspace{0.2in} {\leftline {\bf Acknowledgments}}

We thank Sheldon Katz, Albrecht Klemm, Yuji Sugimoto, Xin Wang, Di Yang for helpful discussions and/or stimulating collaborations on related papers. This work was supported by the national Natural Science Foundation of China (Grants No.11675167 and No.11947301).

\appendix 
\section{Some details on the derivation of the TBA-like equation} \label{appendixA}

Suppose we have a quantum mirror curve of the form
\begin{eqnarray} \label{spectraloperator}
\rho(\hat{x}, \hat{p}) =  u(\hat{x})^{-\frac{1}{2}}   [2\cosh(\hat{p})]^{-1}  u^*(\hat{x})^{-\frac{1}{2}}. 
\end{eqnarray} 
It was conjectured in \cite{Zamolodchikov:1994uw} and proved in \cite{Tracy:1995ax} that the spectral operator is related to some TBA-like equations. We provide some details of the derivations here. We use the variable $\theta \equiv \frac{\pi x}{\hbar}$ which is common in the literature and absorbs the Planck constant. We also define a function $U(x) \equiv  \log(|u(x)|)$, since we can always perform a unitary transformation to absorb the phase in the function $u(x)$, which does not change the spectrum. So the spectrum of the operator (\ref{spectraloperator}) is equivalent to the study of the following integral kernel 
\be
K(\theta, \theta^{\prime}) =\frac{1}{2\pi} \frac{\exp[-\frac{U(\theta)+U(\theta^{\prime})}{2} ]} {2\cosh(\frac{\theta-\theta^{\prime}}{2})}.  
\ee
One introduces  $R_0(\theta)=e^{-U(\theta)}$ and the iterated integral 
\be
R_l(\theta) = e^{-U(\theta)} \int_{-\infty}^{\infty} \frac{\exp[ -\sum_{i=1}^lU(\theta_i) ]}{ \cosh\frac{\theta-\theta_1}{2} 
\cosh\frac{\theta_1-\theta_2}{2}  \cdots \cosh\frac{\theta_l-\theta}{2}  } d\theta_1 \cdots d\theta_l, ~~~ l\geq 1. 
\ee
Furthermore we define a generating series and the decomposition into the even and odd parts 
\be
\ba
& R(\theta| z) =\sum_{l=0}^{\infty} (-\frac{z}{4\pi} )^l R_l(\theta) ,  \\
& R_+(\theta| z) = \frac{1}{2} [ R(\theta| z) + R(\theta| -z)], ~~~  R_-(\theta| z) = \frac{1}{2} [ R(\theta| z) - R(\theta| -z)]. 
\ea
\ee
The expansion parameter $z$ will be identified with the complex structure parameter of the mirror curve. Then the theorems of \cite{Zamolodchikov:1994uw, Tracy:1995ax} state that the even and odd  generating functions can be computed by a TBA system with the integral equations 
\be \label{integralequation}
\ba
U(\theta) &= \epsilon(\theta) +\int_{-\infty}^{\infty} \frac{d\theta^{\prime}}{2\pi} \frac{\log(1+\eta^2(\theta^{\prime}))}{\cosh(\theta-\theta^{\prime})} , \\
\eta(\theta) &=-z \int_{-\infty}^{\infty}  \frac{d\theta^{\prime}}{2\pi} \frac{e^{-\epsilon(\theta^{\prime})} }{\cosh(\theta-\theta^{\prime})}.
\ea
\ee
The TBA integral equations determine the two functions $\eta(\theta), \epsilon(\theta)$ and we have 
\be
\ba
 R_+(\theta| z) = e^{-\epsilon(\theta)}, ~~~~  R_-(\theta| z) = R_+(\theta| z) \int_{-\infty}^{\infty} 
  \frac{d\theta^{\prime}}{\pi} \frac{\arctan(\eta(\theta^{\prime})) }{\cosh^2(\theta-\theta^{\prime})}.
\ea
\ee
To derive the difference equations, we perform Fourier transforms on both sides of the two integral equations (\ref{integralequation}). We use the integral formula $\int_{-\infty}^{\infty}\frac{e^{i \theta \xi} d\theta }{\cosh \theta} = \frac{\pi}{\cosh(\frac{\pi \xi}{2})} $, where $\xi$ is the Fourier conjugate variable of $\theta$. We then multiply both sides by $\cosh(\frac{\pi \xi}{2})$ and perform the inverse Fourier transforms back to the $\theta$ variable,  arriving at the following difference equations  
\be
\ba
U(\theta+\frac{\pi i}{2})+ U(\theta-\frac{\pi i}{2}) &= \epsilon(\theta+\frac{\pi i}{2}) +\epsilon(\theta - \frac{\pi i}{2})  
+\log(1+\eta^2(\theta)), \\
\eta(\theta+\frac{\pi i}{2}) + \eta(\theta - \frac{\pi i}{2}) &=-ze^{-\epsilon(\theta)}  . 
\ea
\ee
We can eliminate the function $\epsilon(\theta)$ and obtain a difference equation for $\eta(\theta)$. In order to compare with quantum periods, we change back the variable $\theta=\frac{\pi x}{\hbar}$ and also need to make some transformations $z\rightarrow z^{-\frac{1}{2}}, \eta\rightarrow i\eta$. We will use the TBA-like difference equation for the function $\eta(X)$ after these manipulations
\begin{eqnarray} \label{TBAgeneral} 
1+z[\eta(qX) +\eta(X)] [ \eta(q^{-1} X) +\eta(X)] |u(q^{\frac{1}{2}} X)u(q^{-\frac{1}{2}} X)| =\eta(X)^2 , 
\end{eqnarray} 
where we use the exponentiated parameters $X=e^x, q=e^{i\hbar}$.

\addcontentsline{toc}{section}{References}


\providecommand{\href}[2]{#2}\begingroup\raggedright\endgroup

\end{document}